# Baryon - baryon correlations in Au+Au collisions at $\sqrt{s_{NN}}$= 62GeV and $\sqrt{s_{NN}}$= 200GeV, measured in the STAR experiment at RHIC.

Hanna Paulina Gos[1,2] for the STAR Collaboration[3]

[1]*SUBATECH*
Laboratoire de Physique Subatomique et des Technologies Associées
EMN-IN2P3/CNRS-Université, Nantes, F-44307, France

[2]*Warsaw University of Technology*
Faculty of Physics, Koszykowa 75, 00-662 Warsaw, Poland
tel. +48 22 234 58 51
fax. +48 22 628 21 71

E-mail: `gos@if.pw.edu.pl`

[3]*http://www.star.bnl.gov/central/collaboration/authorList.php*

**Acknowledgments:** Research supported by the Polish State committee for Scientific Research, grant 2P03B 05925. Research carried out within the scope of the ERG (GDRE): Heavy Ion at ultrarelativistic energies- a European Research Group comprising IN2P3/CNRS, Ecole des Mines de Nantes, Universite de Nantes, Warsaw University of Technology, JINR Dubna, ITEP Moscow and Bogolyubov Institute for Theoretical Physics NAS of Ukraine.

**Abstract**: Particle correlations at small relative velocities can be used to study the space-time evolution of hot and expanding system created in heavy ion collisions. Baryon and antibaryon source sizes extracted from baryon-baryon correlations complement the information deduced from the correlation studies of identical pions. Correlations of non-identical particles are sensitive also to the space-time asymmetry of their emission. High statistics data set of STAR experiment allows us to present the results of baryon-baryon correlation measurements at various centralities and energies, as well as to take carefully into account the particle identification probability and the fraction of primary baryons and antibaryons. Preliminary results show significant contribution of annihilation channel in baryon-antibaryon correlations.

**Keywords:** baryon, correlations, hbt, interferometry, star, rhic.

### *Introduction.*

It is well known that the correlation effects at small relative velocities depend on the distance separating particle emission points in space and time [1,2]. This dependence is reflected by the effects of quantum statistics [3,4] and of the final state interactions [5,6]: Coulomb for charged particles and strong for all types of hadrons. Information about the space-time properties of particle emission is especially important in heavy ion collisions where the complex processes of collective and thermal motion, the rescattering of secondary particles and production of resonances determine the behavior of nuclear matter at the final stage of the collision. [7]

In this context the information coming from the analysis of baryon-baryon correlations can complement the image of the collision emerging from the results of meson-meson correlation studies. The "RHIC HBT puzzle" is still waiting for consistent explanation. Larger masses of baryons with respect to pions or kaons, the ratio of aniproton to proton numbers still less than one at RHIC energies, the differences in the scattering cross sections and the relations with the production of resonanses etc. are the factors showing the reasons for differences between the space-time properties of baryon and meson emission in heavy ion emissions.

In this paper we present the data on baryon-baryon correlations for Au+Au collisions at $\sqrt{(s_{NN})}$= 62GeV and $\sqrt{(s_{NN})}$= 200GeV, measured in the STAR experiment at RHIC.

## *Technical issues.*

Measured particles are registered using the Time Projection Chamber (TPC) of the STAR detector. For both energies the minimum bias events are selected and divided according to percentage of total cross section of Au+Au collisions into three categories: central (0-10%), midcentral (10-30%), peripheral (30-80%) collisions. The two-particle correlations are studied with respect to the momentum of one of the particles in their CMS system, **k\***. Note that the relation between the commonly used variable in the analysis of identical particle correlations Q-invariant is: $Q_{inv}=2k^*$. To measure the correlation effect we define the correlation function as a ratio:

$$C(k^*)=A(k^*)/B(k^*) \qquad (1)$$

We put the pairs of particles coming from the same event (correlated) into numerator *A(k\*)* and the pairs of particles from different events (not correlated) into denominator *B(k\*)*. Protons are measured in the rapidity window *|y|<0.5*. Particles are selected using information from energy loss in the detector (*dE/dx*). Protons and antiprotons are chosen in transverse momentum range: *0.4 GeV/c < $p_T$ < 0.8 GeV/c*. The quality of particle identification is defined by the condition that dependent on the pair type the probability that the analysed pair is composed of proton-proton, antiproton-antiproton, proton-antiproton is at least *0.8*. For combinations of identical particles (proton-proton and antiproton-antiproton) the experimental effect of particle splitting causes artificial enhancement of the number of pairs for low relative momenta $Q_{inv}$. The candidates of splitted tracks are removed from both numerator and denominator of the correlation function. For the combinations of identical and nonidentical particles we remove candidates of merged tracks as well. Assuming that there are no correlations for the pairs with one or two non-primary tracks, the correlation functions are corrected for purity according to the following formula:

$$C_{true}(k^*)=(C_{measured}(k^*)-1)/purity(k^*) + 1, \qquad (2)$$

where $C_{true}(k^*)$ means corrected and $C_{measured}(k^*)$ is measured value of correlation function, *purity(k\*)* is a number describing the probability of correct pair identification, where pair is composed of primary particles.

$$purity(k^*)= Pair\ pid(k^*)^* Fp(p_1)^* Fp(p_2). \qquad (3)$$

*Pair pid(k\*)* corresponds to probability of being registered (dependent on the system) as a: proton-proton, antiproton-antiproton or proton-antiproton pair; *Fp(p$_1$)* and *Fp(p$_2$)* are the probabilities of being a primary track. The significance of purity correction is shown in Fig.1.

The pairs composed of particles with nonidentical masses (protons or antiptotons with lambdas or antilambdas) come from 10% most central Au+Au collisions with $\sqrt{(s_{NN})}$= 200 GeV. Lambdas and antilambdas are reconstructed through the decay channel (e.g. Λ → π⁺ + p). In order to increase the number of proton-lambda pairs the transverse momentum range for protons is 0.4 GeV/c < $p_T$ <1.1 GeV/c and 0.3 GeV/c < $p_T$ <2.0 GeV/c for lambdas and the rapidity interval is |y|<0.5 for protons and |y|<1.5 for lambdas. The mass invariant of the lambda candidates is 1115 ± 6 MeV.

### *Asymmetries in nonidentical particle correlations.*

Nonidentical particle correlation analysis [8] allows to study space-time asymmetries in the emmision of two types of particles (e.g. protons and antiprotons). The distribution of space-time emission points is assumed to be the same for both particles. Proton is always taken as a first in the pair.

The correlation region due to Coulomb interaction and, for baryon systems, nuclear interactions depends on whether two particles move towards each other or away from each other in the pair rest frame. To distinguish two scenarios we calculate $k^*_{out}$ for proton. The case in which $k^*_{out} < 0$ means that proton is slower. If it is emitted further to the center of collision it means that antiproton catches it up and the correlation is stronger for $k^*_{out}<0$. However if it is emitted closer to the center the effect is reversed and the correlation is weaker for $k^*_{out}<0$. Studing the ratio of function for positive and negative values of $k^*_{out}$ the information about space-time asymmetry is deduced as a deviation from the unity. One can study asymmetries in different directions by analysing the correlations dependent on negative and positive components of $k^*$ projections in three directions. We use the decomposition of $k^*$ into out, side and long components.

In STAR due to azimuthal symmetry and symmetry in midrapidity the mean differences:

$$<\Delta r^*_{side}>=<\Delta r^*_{long}>=0 \qquad (4)$$

The asymmetry can be seen for out projection and it is a mixture of two components:

$$<\Delta r^*_{out}>=<\gamma[<\Delta r_{out}>-\beta_T<\Delta t>]> \qquad (5)$$

The correlation functions for different signs of three projections of $k^*$ are presented in Fig.2. Ratios of two functions are equal to unity for all values of $k^*$ for side and long projections confirming proper selection of pairs composed of protons and antiprotons. No deviation from unity in out projections implies no asymmetry in emmision process between protons and antiprotons.

### *Gaussian source approximation.*

The modulus square of pair wave function takes into account quantum statistics and final state interactions (Coulomb and strong).

Proton – proton correlation function for Au+Au collision at $\sqrt{(s_{NN})}$= 62 GeV for different centrality bins is shown in the Fig.3, proton - proton and antiproton – antiproton correlations for Au + Au collision at $\sqrt{(s_{NN})}$= 200 GeV are presented in Fig.4 and Fig.5. For all systems the correlation effect decreases with increasing centrality. The values fit using the CorrFit tool [9] give consistent description within all centralities and energies for all identical systems. We assume the source to be a gaussian.

| Centrality | Radius [fm] | | |
|---|---|---|---|
| | $\sqrt{(s_{NN})}$ = 62 GeV | $\sqrt{(s_{NN})}$ = 200 GeV | |
| | p-p | p-p | $p_{bar}$-$p_{bar}$ |
| minimum bias | $3.1^{+0.2}_{-0.1}$ | $3.4^{+0.1}_{-0.2}$ | $3.4^{+0.2}_{-0.2}$ |
| central: 0-10% | $3.6^{+0.2}_{-0.2}$ | $4.0^{+0.1}_{-0.3}$ | $4.2^{+0.2}_{-0.3}$ |
| midcentral: 10-30% | $3.1^{+0.2}_{-0.2}$ | $3.3^{+0.2}_{-0.1}$ | $3.3^{+0.2}_{-0.2}$ |
| peripheral: 30-80% | $2.3^{+0.2}_{-0.3}$ | $2.5^{+0.2}_{-0.1}$ | $2.3^{+0.3}_{-0.2}$ |

**Table 1.** The centrality dependencies of radii for identical baryon combinations.

The errors presented in the table are statistical ones, however we estimate systematic errors coming from the stability of purity correction to be 0.1fm. Other sources of errors are still under study. We also present the correlation functions for proton-antiproton system for different centrality bins in Fig.6. One can observe similar tendency as for identical combinations: increasing the correlation effect with decreasing the centrality of collision. For all nonidentical systems within all centralities the correlation effect below the unity (annihilation channel) is observed. We also present results of fit, when we again assume the source to be a gaussian.

| Centrality | Radius [fm] | |
|---|---|---|
| | $\sqrt{(s_{NN})}$ = 62 GeV | $\sqrt{(s_{NN})}$ = 200 GeV |
| | p-$p_{bar}$ | p-$p_{bar}$ |
| minimum bias | $2.0^{+0.1}_{-0.1}$ | $2.1^{+0.1}_{-0.1}$ |
| central: 0-10% | $2.3^{+0.1}_{-0.2}$ | $2.5^{+0.2}_{-0.1}$ |
| midcentral: 10-30% | $1.9^{+0.2}_{-0.1}$ | $2.0^{+0.1}_{-0.1}$ |
| peripheral: 30-80% | $1.6^{+0.2}_{-0.2}$ | $1.6^{+0.3}_{-0.2}$ |

**Table 2.** The centrality dependencies of radii for nonidentical baryon combinations.

Presented errors are only statistical ones, but we estimate systematic ones as 0.1fm, taking into account the stability of fit on purity correction. The fitting procedure provides the best theoretical fit taking into account every bin.
We do observe systematically smaller sizes for all systems within all centrality bins for combinations with annihilation channel as compared to the ones without it. Several explanations are possible. Two separate sources with different location of emitting points for different type of particles may exist. In the fitting procedure we neglect the p-wave contribution in all systems which might influence the systems with annihilation channel more, we also did not address the influence of residual correlations. All three reasons still need to be studied in detail.
Lednicky-Lyuboshitz analitical model [6] which can be used for proton-lambda system takes into account strong interaction. It relates the two-particle correlation functions to the scattering amplitude. The function C(k*) is calculated as the square of the wave function $\Psi^S$ averaged over the total spin S and over the assumed gaussian distribution of relative distance **r*** (particle emission points in the pair rest frame):

$$C(k^*) = \langle |\Psi^S_{-k^*}(r^*)|^2 \rangle \qquad (6)$$

The approximated wave function takes into account only the s-wave part of the scattered wave:

$$\Psi^S_{-k^*}(r^*) = e^{-k^* r^*} + f^S(k^*) e^{ik^* r^*}/r^*. \qquad (7)$$

The effective range approximation for the s-wave approximation is given by the formula:

$$f_S(k^*) = [(f_S^0)^{-1} + 0.5\, d_S^0\, k^{*2} - ik^*]^{-1} \qquad (8)$$

where $f_S^0$ is scattering length and $d_S^0$ is the effective radius for a given total spin S=0 or S=1 for triplet (t) or singles (s) state respectively. The correlation function is calculated analytically for a gaussian r* distribution:

$$d^3N/d^3r^* \sim \exp(-r^{*2}/4r_0^2) \qquad (9)$$

where $r_0$ is the radius of source.

We obtain two different radii not only for systems with same masses, this effect also exists in strange baryon correlations, in systems composed of protons and lambdas.

We present the nonidentical, strange correlation functions in the Fig.7. Once again the correlation effect is wider for systems with annihilation channel (proton-antilambda and antiproton-lambda) comparing to proton-lambda and antiproton-antilambda [10]. The radii are:

| System | $r_0$ [fm] |
|---|---|
| p$_{bar}$ − Λ + p − Λ$_{bar}$ | $1.50 \pm 0.05^{+0.10}_{-0.12}$ |
| p − Λ + p$_{bar}$ − Λ$_{bar}$ | $3.09 \pm 0.30^{+0.17}_{-0.25}$ |

**Table 3**. The results of radii for strange baryon correlations.

*Summary.*
In this paper we present baryon-baryon correlations measured by the STAR experiment. Huge amount of data for Au+Au collisions collected recently allows to measure with such high precision proton-proton correlations for different centralities and energies of collision. The antiproton-antiproton correlation is measured for the first time ever. Moreover, for the first time we present measured and fitted proton- antiproton correlation functions. All identical systems within all centrality bins and energies give consistent results and complement ones obtained for identical pion interferometry. Nonidentical baryon correlations of systems with identical masses create consistent description as well, however comparing all combinations together we record smaller radii for all systems with annihilation channel. These studies confirm similar phenomena seen before in strange baryon correlations. We do plan in the future to analyse two- and three-dimensional proton-proton correlations, as the statistics collected by the STAR detector is sufficient to calculate it. More detailed study is required in order to learn more about physics in baryon systems.

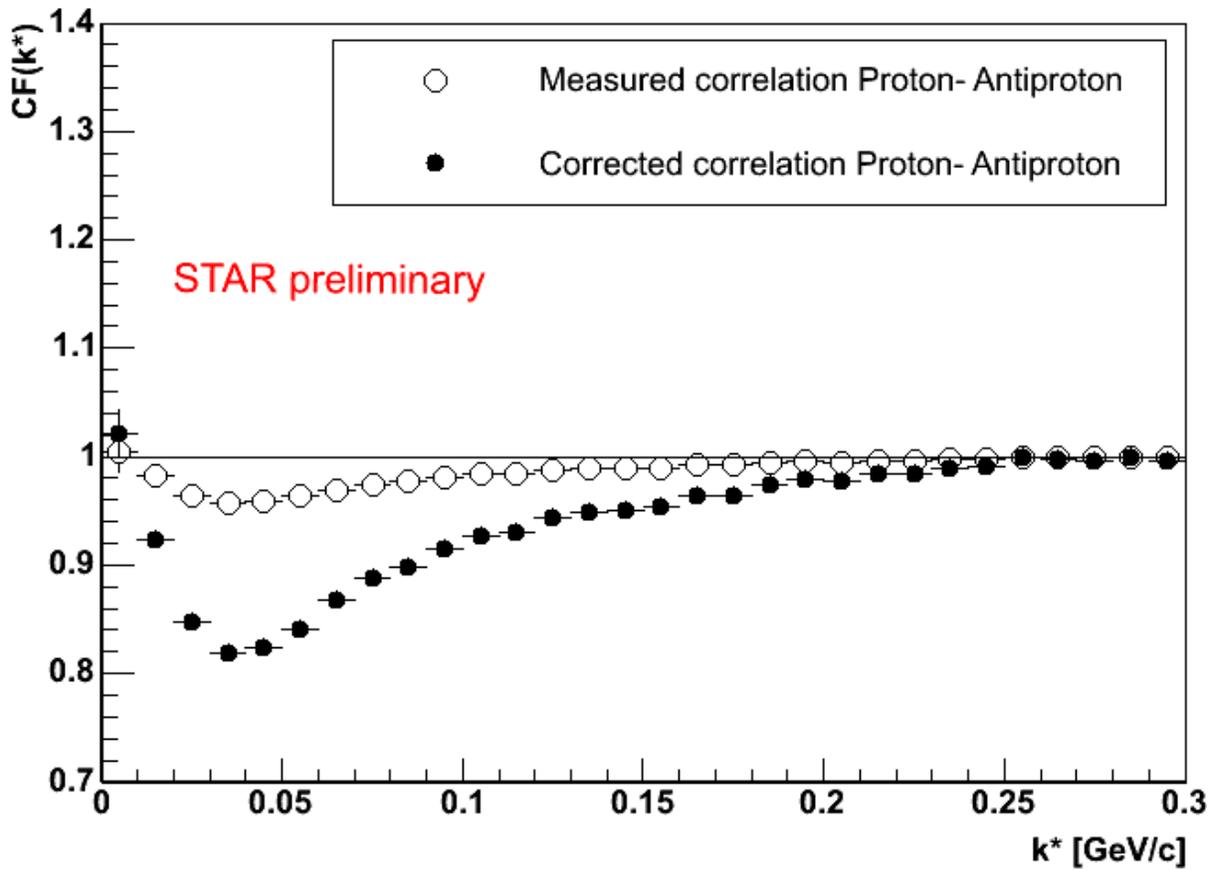

Fig1. The comparison of results for proton-antiproton system corrected and not-corrected for purity shows how significant is precise application of purity correction to proper understanding of the results.

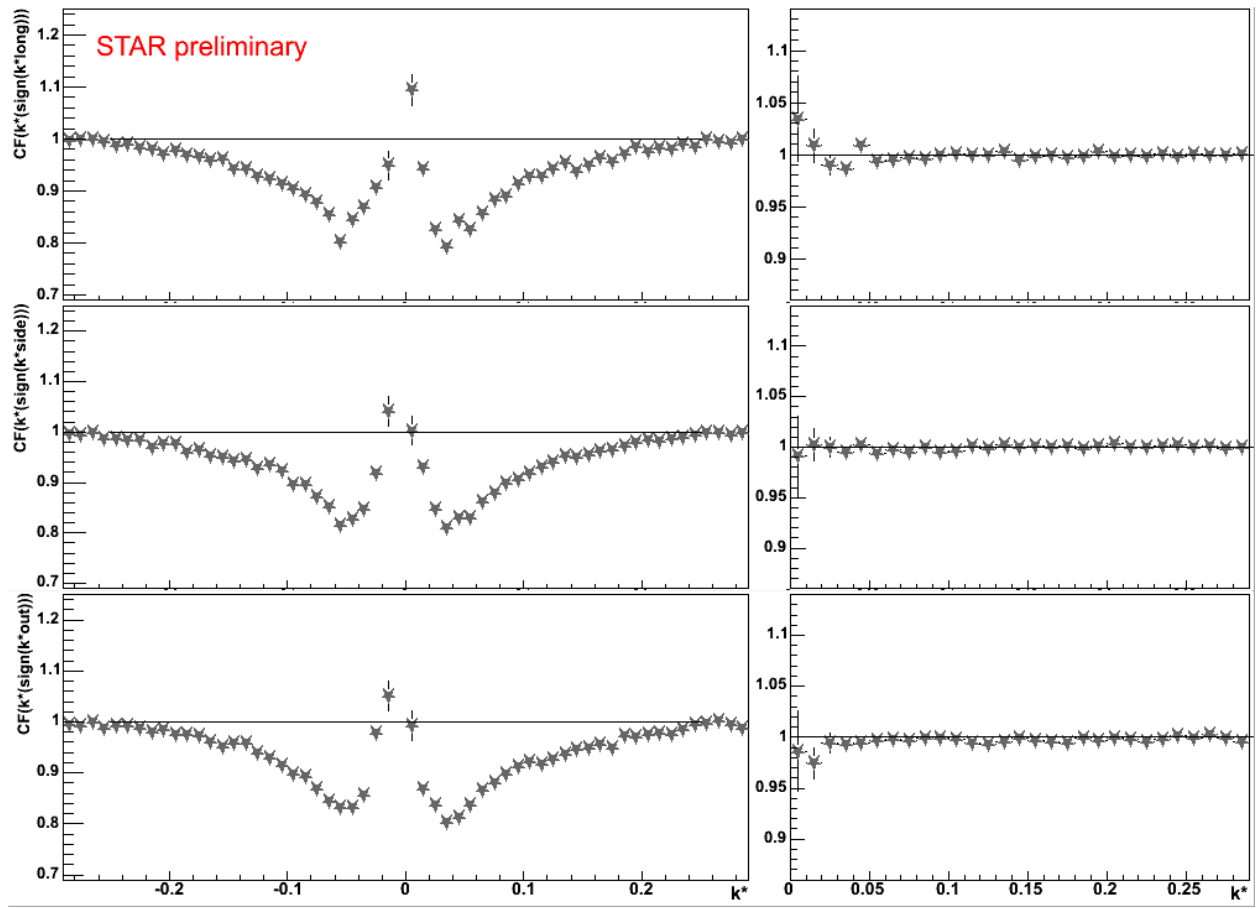

Fig2. Proton- antiproton correlation functions for projections: (from top to bottom): Long, Side, Out and their double ratios.

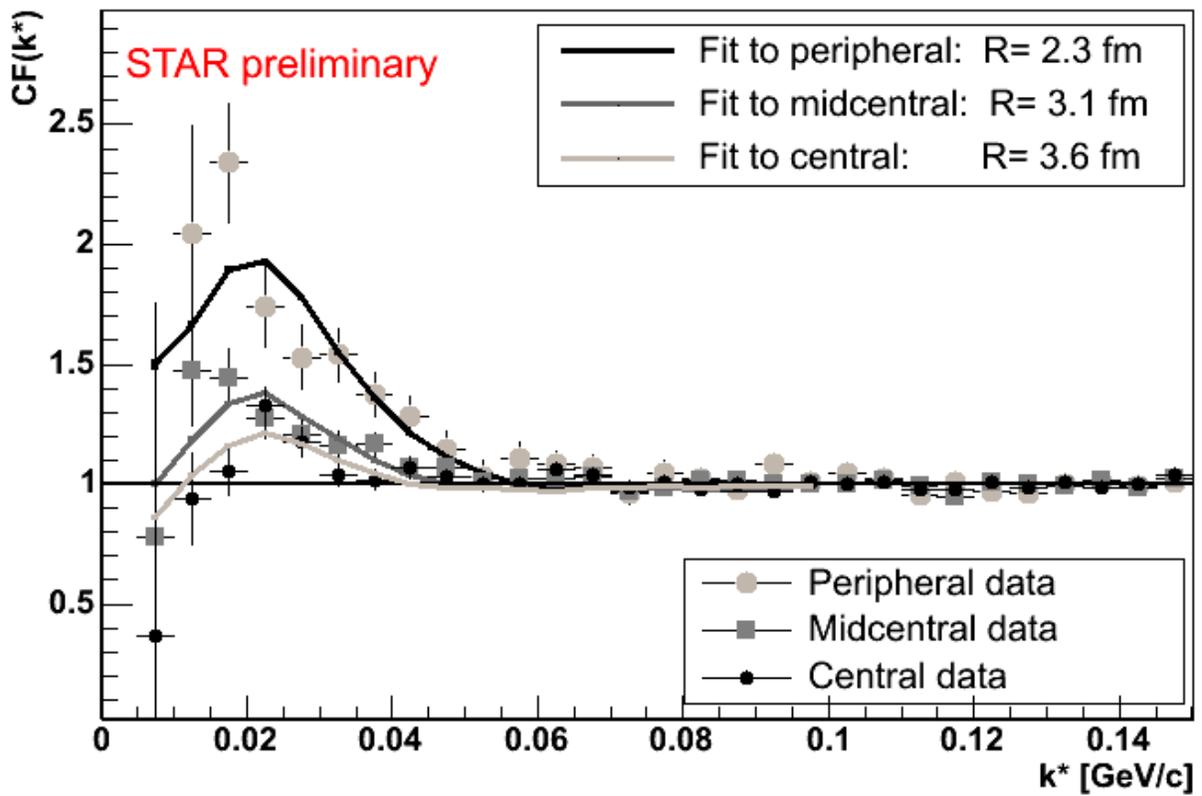

Fig.3. Proton- proton correlation function for various centrality bins for $\sqrt{s_{NN}}$ = 62 GeV

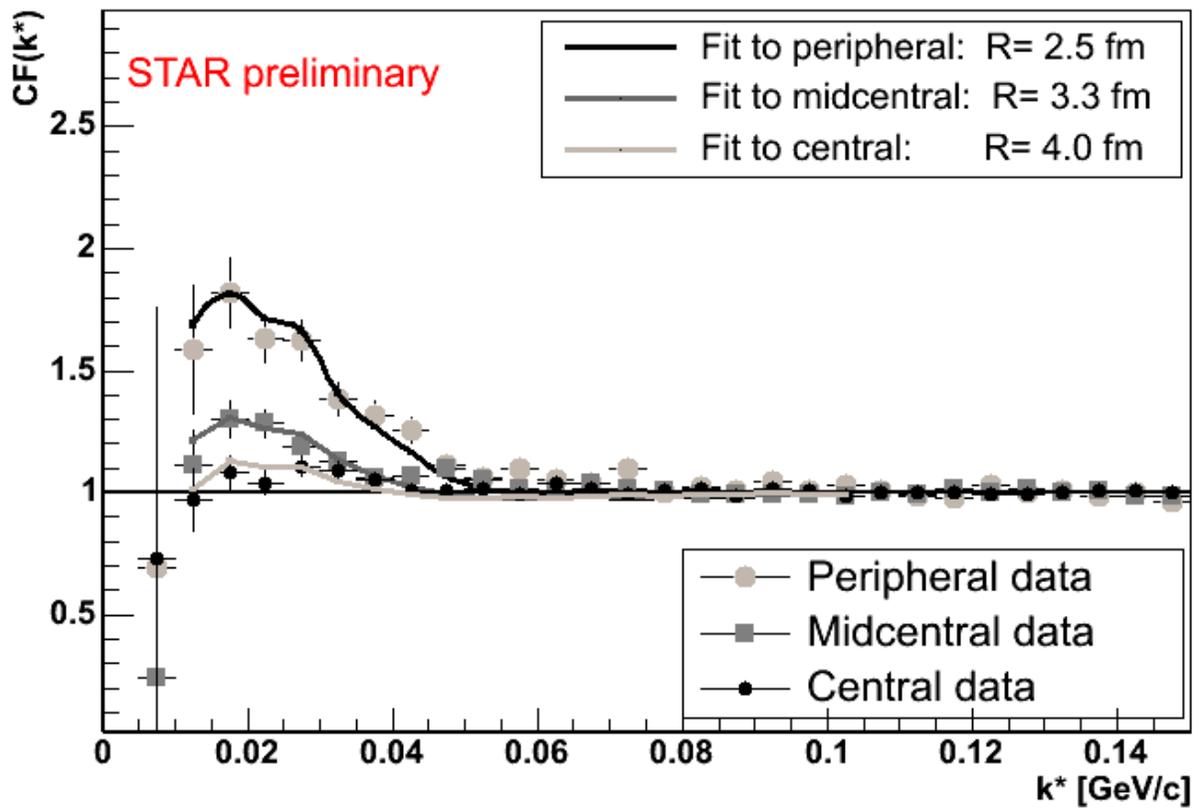

Fig.4. Proton- proton correlation function for various centrality bins for $\sqrt{s_{NN}}$ = 200 GeV.

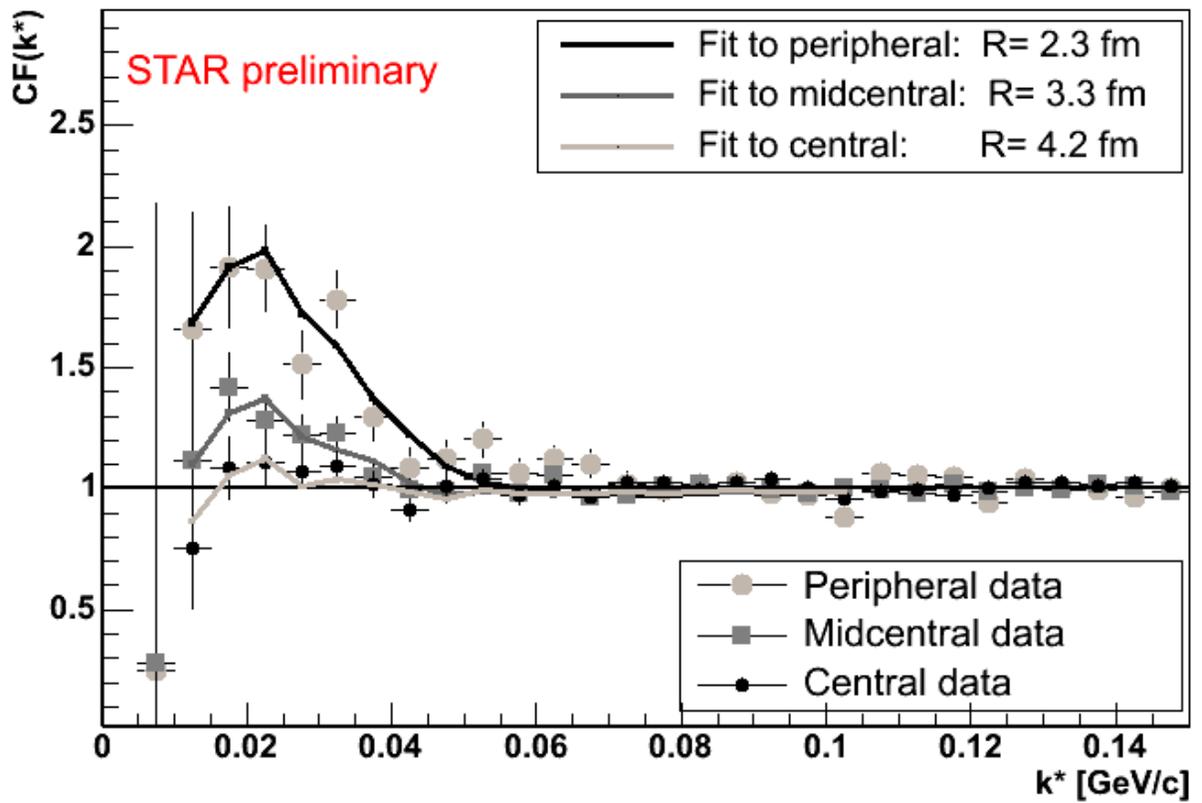

Fig.5. Antiproton- antiproton correlation function for various centrality bins for $\sqrt{(s_{NN})}$ = 200 GeV

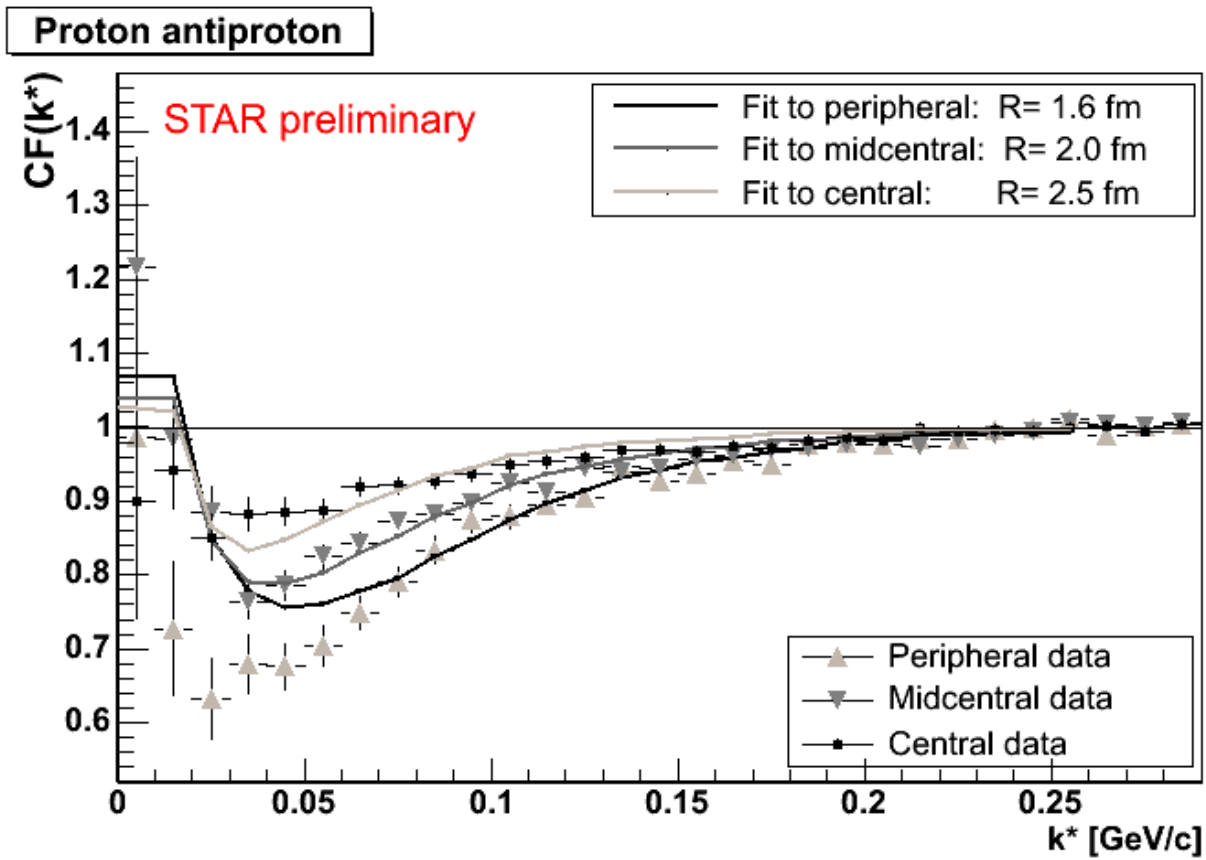

Fig.6. For nonidentical systems annihilation channel is significant.
The figure illustrates centrality dependencies for antiproton – antiproton correlations for $\sqrt{s_{NN}} = 200$ GeV.

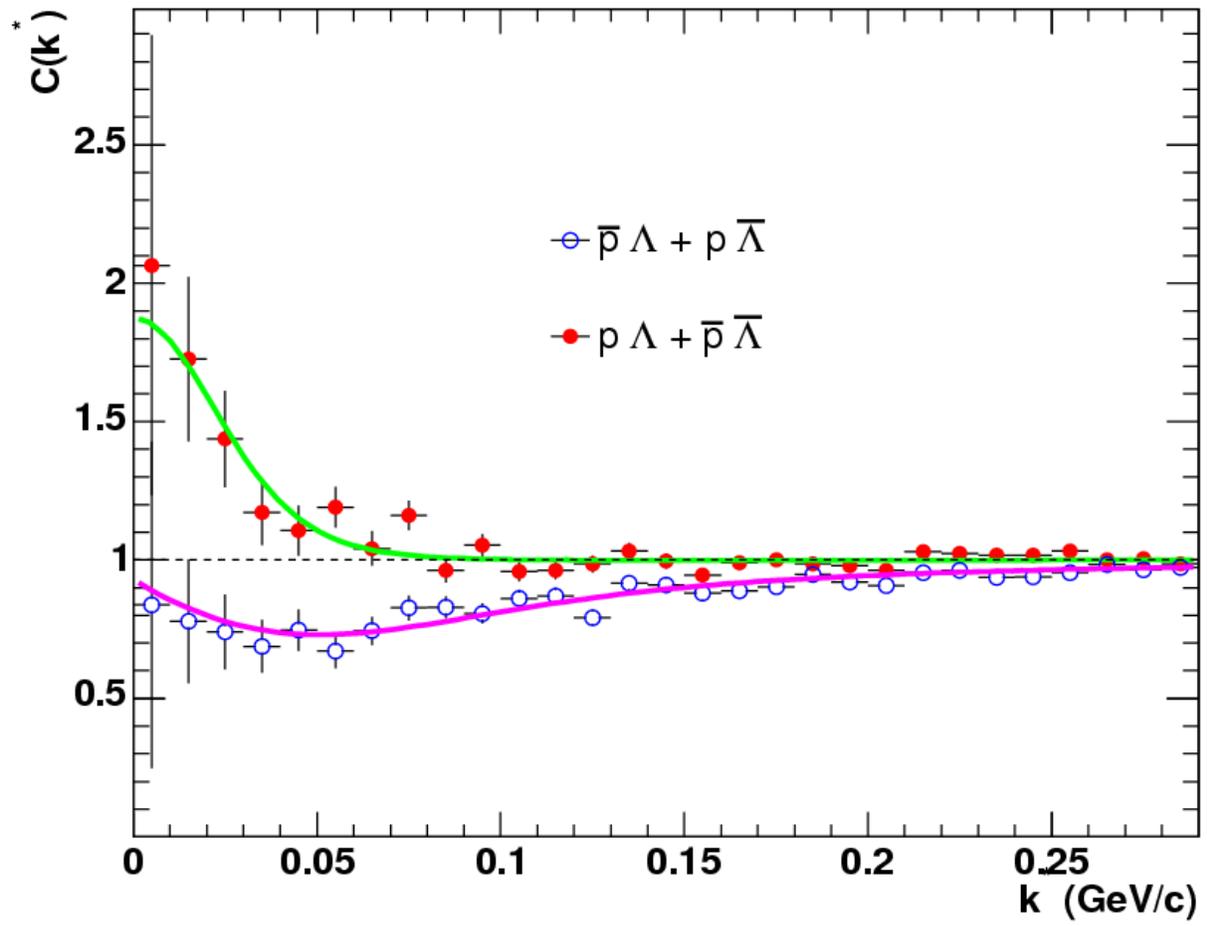

Fig.7. Combined proton- lambda correlation functions. All are corrected for purity.